# Pressure induced electronic topological transition in $Sb_2S_3$

Y. A. Sorb[1], V. Rajaji[1], P. S. Malavi[2], U. Subbarao[3], P. Hadappa[3], S. Karmakar[2], Sebastian C. Peter[3] and C. Narayana[1]*

[1]Chemistry and Physics of Materials Unit, JNCASR, Jakkur, Bangalore 560 064, India
[2]HP&SRPD, Bhabha Atomic Research Centre, Trombay, Mumbai 400085, India
[3]New Chemistry Unit, JNCASR, Jakkur, Bangalore 560 064, India

## Abstract

Pressure induced electronic topological transitions in the wide band gap semiconductor $Sb_2S_3$ ($E_g$ = 1.7-1.8 eV) with similar crystal symmetry (SG: *Pnma*) to its illustrious analog, $Sb_2Se_3$, has been studied using Raman spectroscopy, resistivity and the available literature on the x-ray diffraction studies. In this report, the vibrational and the transport properties of $Sb_2S_3$ have been studied up to 22 GPa and 11 GPa, respectively. We observed the softening of phonon modes $A_g(2)$, $A_g(3)$ and $B_{2g}$ and a sharp anomaly in their line widths at 4 GPa. The resistivity studies also shows an anomaly around this pressure. The changes in resistivity as well as Raman line widths can be ascribed to the changes in the topology of the Fermi surface which induces the electron-phonon and the strong phonon-phonon coupling, indicating a clear evidence of the electronic topological transition (ETT) in $Sb_2S_3$. The pressure dependence of a/c ratio plot obtained from the literature showed a minimum at ~ 5 GPa, which is consistent with our high pressure Raman and resistivity results. Finally, we give the plausible reasons for the non-existence of a non-trivial topological state in $Sb_2S_3$ at high pressures.

## I. INTRODUCTION

New electronic states of matter known as topological insulators have received a wide spread attention due to their potential applications in spintronics and quantum computing devices and have rejuvenated the field of condensed matter physics in the last few years.[1-4] Unlike conventional insulators, topological insulators have gapless spin-filtered edge electronic states on their surfaces on account of strong spin-orbit coupling, which permit the dissipation less transport of charge and spin; and these states are protected by time-reversal symmetry.[5-8] The metal chalcogenide semiconductors belonging to the $A_2B_3$ (A= Bi, Sb, As; B=S, Se, Te) series has been of great interest owing to their exceptional thermoelectric properties and technological applications.[9-11] Today the quest of topological insulators has widened beyond $A_2B_3$ chalcogenide series and many compounds have been theoretically proposed and experimentally realized as topological insulators.[12-15] For instance, $Bi_2Se_3$, $Bi_2Te_3$ and $Sb_2Te_3$ compounds in the $A_2B_3$ series have single Dirac cone at the Γ point of the Brillouin zone and they adopt Rhombohedral structure (S.G: $R\bar{3}m$, Z=3) at ambient pressure. These compounds are extensively studied 3D topological insulators at ambient pressure.[16-20] Upon pressurizing them, they exhibit novel phenomena such as structural phase transitions,[21-23] electronic topological transitions (ETT) or Lifshitz transition,[21-25] insulator to metal transitions[26] and superconductivity[26-27]. The topological insulators $Bi_2Te_3$ and $Sb_2Te_3$ undergo series of pressure induced structural phase transitions from Rhombohedral ($R\bar{3}m$) to sevenfold monoclinic (S.G: $C2/m$, Z=4) followed by eightfold monoclinic ($C2/c$, Z=4) and finally to disordered body-centered cubic structure (space group $Im\text{-}3m$, Z=2), respectively.[21-22] In addition to this, $Sb_2Te_3$ adopt a bcc like monoclinic phase ($C2/m$) at an intermediate pressure. On the contrary, high pressure structural behavior of $Bi_2Se_3$ is more complex.[23]

Using first principle calculations based on density functional theory, Koushik Pal *et al.* predicted that *β*-$As_2Te_3$ (S.G: $R\bar{3}m$) can be transformed from band insulator to a non trivial topological insulator under uniaxial strain of ~1.77 GPa.[28] Another class of compounds (wide band gap semiconductors) formed in the $A_2B_3$ series are $Sb_2Se_3$ [$E_g$=1 eV], $Bi_2S_3$ [$E_g$=1.3 eV] and $Sb_2S_3$ [$E_g$=1.7-1.8 eV];[29] unlike $Bi_2Te_3$, $Sb_2Te_3$ and $Bi_2Se_3$ compounds, these compounds adopt orthorhombic structure (SG: *Pnma*, Z=4, $U_2S_3$-type).[24,30] Among these three wide band gap semiconductors, $Sb_2Se_3$ has recently been received wide attention in the view point of its topological behavior at high pressure.[26, 31-32] Recent high pressure study on this compound by Kong *et al.* showed a pressure induced superconductivity at above 10 GPa preceeded by an insulator to metal like transition at ~3 GPa and they found that this transition could be related to the topological quantum phase transition (TPQT).[26] Moreover, Efthimiopoulos *et al.* observed structural transformation of $Sb_2Se_3$ to a disordered cubic bcc [S.G: *Im-3m*] alloy at above 51 GPa using high pressure X-ray diffraction study.[30] Similarly, high pressure structural and vibrational properties of $Bi_2S_3$ up to ~65 GPa have been reported. In this study, the ambient *Pnma* structure was found to persist up to 50 GPa and suffered from structural disorder upon pressurizing further.[24] An important finding in this study was the possibility of ETT at around 4-6 GPa due to the topological modification of the $Bi_2S_3$ electronic structure.[24] A significant characteristic feature of the compounds in the $A_2B_3$ chalcogenide series is the existence of a second order isostructural phase transition called ETT below 6 GPa.[21-26] Among the metal chalcogenide compounds in the $A_2B_3$ series, $Sb_2S_3$ has not received much attention as compared to other compounds, which may probably be due to its wide band gap, that causes lack of topological properties. It has good photovoltaic properties, high thermoelectric power and broad spectrum response.[33] It has been used in various application such as television cameras with photo conducting targets, thermoelectric cooling devices, electronic and optoelectronic devices, solar energy conversion and visible light-responsive photo catalysis.[33] The only high pressure study reported on this compound to date is the X-ray diffraction study up to 10 GPa and no structural phase transition is noticed up to 10 GPa.[34] In this report, we present our experimental observation of ETT in $Sb_2S_3$ for the first time using high pressure Raman and resistivity studies.

## II. EXPERIMENTAL DETAILS

The pure phase of $Sb_2S_3$ was synthesized by combining 0.7168 g of antimony in the form of shots (99.99%, Alfa Aesar) and 0.2832 g of sulfur in powder form (99.99% Alfa Aesar), were placed in a 9 mm diameter quartz tube, under an inert (argon) atmosphere inside a glove box, which was flame-sealed under vacuum of $10^{-3}$ torr, to prevent oxidation during heating. The tube was then placed in a vertically aligned tube furnace and heated to 120 $^oC$ over the period of 2 h to allow proper homogenization. Subsequently, the temperature was increased to 600 $^oC$ and kept for 2 days. Finally, the system was allowed to cool to room temperature in 10 h. No reaction with the quartz tube was observed. A light grey polycrystalline $Sb_2S_3$ was formed and it was found to be stable in moist air for several months. The weight losses of the final material were found to be less than 1%. Phase identity and purity of the sample was determined by powder X-ray diffraction study (XRD), those were carried out with a Bruker D8 Discover diffractometer using Cu-$K\alpha$ radiation ($\lambda$ = 1.5406 Å) over the angular range $20^0 \leq 2\theta \leq 80^0$, at room temperature, calibrated against corundum standard. The experimental powder patterns of $Sb_2S_3$ and the XRD pattern simulated from the reported data were found to be in good agreement.[35] The sample was then used for high pressure Raman and resistivity studies. High pressure Raman studies were carried out using a Mao-Bell type Diamond Anvil Cell (DAC). A stainless steel gasket was preindented to a thickness of ~60 µm and a hole of diameter ~200 µm was drilled at the centre of the gasket which was used as sample chamber. The fine powdered sample was loaded into the sample chamber, in which methanol and ethanol in 4:1 ratio was used as pressure transmitting medium (PTM). Pressure was determined by using ruby fluorescence technique. Raman measurements were carried out by a custom-built micro-Raman spectrometer with a laser excitation of 532 nm using a frequency-doubled Nd-YAG laser with power of 4 mW and the scattered light was collected in a backscattering geometry using a CCD detector.[36] The pressure dependence of resistivity at room temperature has been measured on $Sb_2S_3$ up to ~10 GPa using a miniature DAC and an optical cryostat. A quasi-four probe resistance measurement technique with *in-situ* ruby pressure measurement was employed. The details of the DAC preparation, resistance measurement and sample pressure determination technique have been described elsewhere.[37] Resistance was measured on a pressed pellet of $Sb_2S_3$ (diameter ~50 µm and ~20µm thick) using 10 µm thin gold electrical leads, that remains in good contact with the sample by the applied pressure through the solid PTM. Finely powdered NaCl (soft solid, remains insulating throughout this temperature and pressure range) was used as the PTM.

### III. RESULTS AND DISCUSSION

The orthorhombic (S.G: *Pnma*, point group symmetry, $D_{2h}$) unit cell of the $Sb_2S_3$ consists of 20 atoms, which have 60 phonon active modes at the Γ point of the Brillouin zone.[38] Since *Pnma* is a centrosymmetric space group, both Raman and infrared (IR) modes of $Sb_2S_3$ are mutually exclusive.[38-39]

$$\Gamma_{vib} = 30\ \Gamma_{Raman} + 22\ \Gamma_{IR} + 3\ \Gamma_{acoustic} + 5\ \Gamma_{silent}$$

Where $\Gamma_{acoustic} = B_{1u} + B_{2u} + B_{3u}$ are acoustic modes, $\Gamma_{Raman} = 10\ A_g + 5\ B_{1g} + 10\ B_{2g} + 5 B_{3g}$ are Raman active modes, $\Gamma_{IR} = 4\ B_{1u} + 9\ B_{2u} + 9\ B_{3u}$ are IR active modes and $\Gamma_{silent} = 5 A_u$ are silent modes.

In $Sb_2S_3$ lattice, all the five (two Sb and three S) atoms are situated on Wyckoff positions 4c, at the site symmetry $C_s$ with four fold multiplicity. The motion of Sb and S atoms are within xz plane for the $B_{2g}$, $B_{1u}$ and $B_{3u}$ modes and it is along y-axis for $B_{1g}$, $B_{3g}$ and $B_{2u}$ modes.[38] Raman measurements on $Sb_2S_3$ in our present study shows that there are six Raman active modes present at ambient pressure. Nevertheless, there is no systematic experimental study available to assign these phonon modes.

Sereni *et al.* carried out polarization dependent Raman scattering experiments on $Sb_2S_3$, in which they have recorded backscattering geometry with parallel polarization $A_g$ $x(yy)\bar{x}$ and 90° geometry with parallel polarization $A_g$ $\bar{x}(yy)z$ and they observed the occurrence of $B_{1g}$ $\bar{x}(yx)z$ spectrum in their $A_g$ spectrum.[39] However, the authors couldn't distinguish between $A_g$ and $B_{2g}$ modes and also that of $B_{1g}$ and $B_{3g}$ owing to the lack of informations about the intensities of phonon modes. For the present study, we could resolve six Raman active modes at ambient pressure (300 K) as shown in lower part of Fig. 1. The peaks at 195 $cm^{-1}$, 208 $cm^{-1}$ and 241 $cm^{-1}$ appear to be very weak due to poor background, but, the spectra was well resolved at low temperature (77K) as shown in upper part of Fig. 1. The experimentally observed phonon modes of $Sb_2S_3$ has been assigned based on the lattice dynamics calculation by Liu *et al.* and the observed phonon mode frequencies are shown in Table 1.[38] In Fig.1, the phonon modes at 77 K is softened with respect to the Raman spectra collected at 300 K, which is due to the compression of lattice upon reducing the temperature. For our convenience, we represent the three different $A_g$ modes by a superscript which numbers the modes in terms of increasing frequency. High pressure Raman studies has been carried out on $Sb_2S_3$ in a hydrostatic environment up to 22 GPa and the representative Raman spectra of $Sb_2S_3$ at various pressures are shown in Fig.2. It is evident from Fig.2, apart from the observed phonon modes; the appearance of additional Raman modes above 20 GPa suggests the compound may be undergoing structural phase transition beyond 20 GPa. Eventhough, both $Sb_2Se_3$ and $Bi_2S_3$ compounds adopt identical crystal structure with $Sb_2S_3$, no structural phase transition below 50 GPa is reported from these compounds.[4,30] However, high pressure X-ray diffraction studies are required to confirm our observation. In this study, our interest is focused only on the pressure induced ETT in $Sb_2S_3$ rather than its structural transitions under high pressure. All the phonon modes are fitted with Lorenzian line shape function. Due to the poor signal to noise ratio, line width analysis of $A_g^1$, $B_{1g}$ and $B_{3g}$ modes are highly unreliable. Hence, out of six experimentally observed Raman active modes, we could able to fit only three high intense modes namely $A_g^2$, $A_g^3$ and $B_{2g}$ at various pressures. Pressure dependence of $A_g^2$ (286 cm$^{-1}$), $A_g^3$ (305 cm$^{-1}$) and $B_{2g}$ (314 cm$^{-1}$) phonon modes are shown in Fig.3. The softening of phonon modes are noticed at low pressure regions. The intensity of $B_{2g}$ (314 cm$^{-1}$) mode decreases with pressure and at around 4 GPa, it is either disappeared or completely merged with the $A_g^3$ (305 cm$^{-1}$) mode. The frequency of $A_g^3$ (305 cm$^{-1}$) mode is softened below 4 GPa and a discontinuity in frequency was seen at 4 GPa; afterward, the mode is again softened upon pressurizing further. The behavior of $A_g^2$ mode is quite different as compared to the other modes. As is evident from Fig.3, the $A_g^2$ mode is softened with pressure and reaches a minimum at around 4 GPa and thereafter the mode is hardened upon pressurizing further. The softening and the disappearance of $B_{2g}$ (314 cm$^{-1}$) mode, a discontinuity in frequency of $A_g^3$ (305 cm$^{-1}$) mode and moreover the hardening of $A_g^2$ phonon mode above 4 GPa hint a phase transition at 4 GPa and the transition thus noticed is reproducible. A study of phonon life time change with pressure is useful to get more insight into the phonon anomalies during the transition. The pressure dependence of the full width at half maximum (FWHM) of $A_g^2$, $A_g^3$ and $B_{2g}$ modes of $Sb_2S_3$ as shown in Fig.4. It is evident from Fig.4 that the line width of the $B_{2g}$ (314 cm$^{-1}$) mode increases substantially with pressure up to 4 GPa and afterward it disappears. A sharp anomaly in the line width of $A_g^3$ (305 cm$^{-1}$) mode is noticed at ~ 4 GPa. The line width of $A_g^2$ mode is almost independent of pressure up to 4 GPa and thereafter it decreases with pressure. The large line width of $Sb_2S_3$ phonon modes can be explained by assuming the anharmonic decay into two phonons of lower frequency as a result of strong phonon-phonon interaction, which is possible, when the frequencies of the first order phonon modes coincide with the high density of two-phonon density of states.[40,23] The large line width and the vanishing (or substantial reduction in intensity) of $B_{2g}$ mode and also the decrease in the line width of $A_g^3$ phonon mode are indications of huge changes in the topology of Fermi surface as a result of strong phonon-phonon coupling. The line width of the phonon modes is correlated to the position of the phonon frequency and the van Hove singularities associated to the two phonon density of states.[40]

High pressure X-ray diffraction studies reported on $Sb_2S_3$ showed that there is no struc-tural phase transition up to 10 GPa [50], which suggests the observed anomalies of the phonon modes and their line widths at 4 GPa is not due to structural phase transition. In contrast, the observed changes can be attributed to an isostructural phase transition. Further, an interchange of lattice parameters "a" and "c" at around 1.2 GPa was observed by Lundegaard *et al.*, which motivated us to plot pressure dependence of a/c ratio and surprisingly, we noticed a minimum at around 5 GPa from P Vs a/c as shown in Fig. 5.[34] The asymmetric behavior of P Vs a/c ratio is already well documented in other compounds of metal chalcogenide $A_2B_3$ series and was attributed to an ETT.[22-26] An ETT is an isostructural second order phase transition, no volume discontinuity and no change in Wyckoff positions are anticipated during this transition.[41] An ETT causes anomalous signature in mechanical, electrical and thermodynamic properties and it also affects vibrational properties.[41-42] Hence, we interpret the anomaly at around 4 GPa in $Sb_2S_3$ can be due to a second order isostructural phase transition called ETT in comparison with similar transition observed in the metal chalcogenide $A_2B_3$ series.[21-26]

The electrical transport measurement is a versatile tool to detect the subtle changes in the topology of the Fermi surface during ETT. Hence, we carried out electrical transport measurements of this compound under high pressure. The pressure dependent electrical resistivity (ρ) of $Sb_2S_3$ at room temperature is shown in Fig.6. A non-monotonic increase of resistivity is observed below 1.4 GPa and two points of inflections are noticed at 1.4 GPa (maximum) and 2.4 GPa (minimum) respectively. Thereafter, the resistivity increases monotonically with pressure up to 4 GPa and it is almost saturated on further pressurizing the sample. Albeit, high pressure resistivity study shows a sharp maximum and minimum at 1.4 GPa and 2.4 GPa, we couldn't observe any anomalies in both high pressure Raman and XRD studies in the respective pressure regions. At present, we are not sure about the cause of the changes observed in the resistivity at 1.4 GPa and 2.4 GPa and detailed band structure calculation is required to understand the origin of these changes at the respective pressures. Additionally, at 4 GPa an obvious change in the slope of the resistivity is observed which can be attributed to an ETT. The weak signature of resistivity at 4 GPa may probably be due to the large resistivity value of the system. This sort of subtle changes in the resistivity value at high pressure has already been documented in $Bi_2Te_3$ topological insulator.[43] The change in slope of the pressure Vs resistivity plot at 4 GPa supports the presence of anomaly of electronic states during ETT, which can possibly be due to electron-phonon coupling. The ETT reported to date in the metal chalcogenide $A_2B_3$ series as shown in the Table 2. For decades before, ETT was studied by Lifshitz and he pointed out that during such a transition, electronic band structure extremum associated to a Van Hove singularity in the density of states crossing the Fermi energy causes a change in the topology of the Fermi surface.[44]

It is quite interesting to compare why $Sb_2Se_3$ is a probable non trivial topological insulator and $Sb_2S_3$ a band insulator at high pressure eventhough both of them have same crystal symmetry. The reason is that the compounds formed by elements with atomic number greater than 50 are naturally among the prime candidates of topological insulators as a result of strong spin orbit coupling, which must be strong enough to have considerable effect to modify electronic structure. It is well known that the spin-orbit coupling strength is proportional to $Z^4$ (where $Z$ is the atomic number; $Z = 51$ for Sb, 34 for Se and 16 for S).[31] Consequently, the spin-orbit coupling is not strong enough to generate the band inversion in both $Sb_2Se_3$ and $Sb_2S_3$ compounds at ambient pressure. However, these band insulators can probably be transformed in to a nontrivial topological insulator by applying pressure owing to the enhancement of the crystal-field splitting, resulting a crossover of conduction and valence bands which leads to TQPT.[31] Since the band gap of $Sb_2S_3$ ($E_g$ = 1.7-1.8 eV) is quite large as compared to the energy scale of spin-orbit coupling, therefore the spin-orbit coupling is not strong enough to change the quantum phase transition.[3] High pressure resistivity results suggests that $Sb_2S_3$ is still in trivial insulating state at 4 GPa. The recent theoretical and subsequent experimental investigations insist that $Sb_2Se_3$ is a pressure induced topological insulator.

## IV. CONCLUSION

High pressure Raman and electrical resistivity studies were used to investigate the vibrational and transport properties of $Sb_2S_3$. Sharp changes observed in the phonon frequencies $A_g^2$, $A_g^3$ and $B_{2g}$ and their line widths at 4 GPa can be stemmed from the strong phonon-phonon coupling. Electrical transport measurements carried out under high pressure showed a change in slope at 4 GPa could possibly be due to the electron-phonon coupling. Also, a minimum at around 5 GPa is observed by plotting P Vs a/c ratio obtained from the literature. From all these evidences (phonon anomalies, change in resistivity and a minimum in a/c ratio), we attributed the observed transition at 4 GPa is an ETT. Finally, the possible reasons for the absence of pressure induced topological non trivial state of $Sb_2S_3$ is discussed.

## ACKNOWLEDGMENTS

We acknowledge springer (License Number: 3553701423319) for granting permission to use high pressure XRD data of $Sb_2S_3$ from Phys. Chem. Minerals. We gratefully acknowledge Prof. Alfredo Segura, University of Valencia for valuable discussions. Y. A. Sorb acknowledges JNCASR for granting post doctoral fellowship V. Rajaji acknowledges for the financial support from DST and JNCASR, USR thanks CSIR for the research fellowship and SCP thanks DST for the Ramanuajan fellowship.

**TABLE I.** Mode assignment of experimentally observed Raman spectra of $Sb_2S_3$ at ambient pressure

| Raman shift ($cm^{-1}$) | Raman modes |
|:---:|:---:|
| 195 | $A_g^1$ |
| 208 | $B_{1g}$ |
| 241 | $B_{3g}$ |
| 286 | $A_g^2$ |
| 305 | $A_g^3$ |
| 314 | $B_{2g}$ |

**TABLE II.** Comparison of pressure induced ETT of different metal chalcogenides in $A_2B_3$ series.

| Compound | $E_g$ (eV) | Space group | Transition pressures (GPa) |
|---|---|---|---|
| $Bi_2Se_3$ | 0.30 | $R\bar{3}m$ | 3-5[a] |
| $Bi_2Te_3$ | 0.12 | $R\bar{3}m$ | 3.2 or 4[b] |
| $Sb_2Te_3$ | 0.28 | $R\bar{3}m$ | 3-3.5[c] |
| $Sb_2Se_3$ | 1.00 | *Pnma* | 2.5[d] |
| $Bi_2S_3$ | 1.30 | *Pnma* | 4[e] |
| $Sb_2S_3$ | 1.7-1.8 | *Pnma* | 4[f] |

[a]Reference [23].
[b]Reference [21,25].
[c]Reference [22].
[d]Reference [26].
[e]Reference [24].
[f]Reference [present work].

**Figure Captions**

**Fig. 1.** Raman spectra of $Sb_2S_3$. The upper and lower parts represent the Raman spectra collected at 77 K and 300 K respectively.

**Fig. 2.** Raman spectra of $Sb_2S_3$ at various pressures. The observed phonon modes of $Sb_2S_3$ are assigned at ambient pressure ($10^{-4}$ GPa).

**Fig. 3.** P Vs Raman shift of $A_g^2$, $A_g^3$ and $B_{2g}$ modes of $Sb_2S_3$. A vertical dotted line at 4 GPa is to indicate an ETT. The dotted lines drawn to each phonon modes are for guidance to the eye.

**Fig. 4.** P Vs FWHM of $A_g^2$, $A_g^3$ and $B_{2g}$ modes of $Sb_2S_3$. A vertical dotted line at 4 GPa is to indicate an ETT. The dotted lines drawn to each phonon modes are for guidance to the eye.

**Fig. 5.** P Vs a/c ratio of $Sb_2S_3$ up to 10 GPa [Adapted from L. F. Lundegaard *et al.*].

**Fig. 6.** P Vs electrical resistivity of $Sb_2S_3$. A vertical dashed red line at 4 GPa is to indicate the ETT. The dotted lines drawn to the resistivity data is for guidance to the eye.

**Fig. 1.**

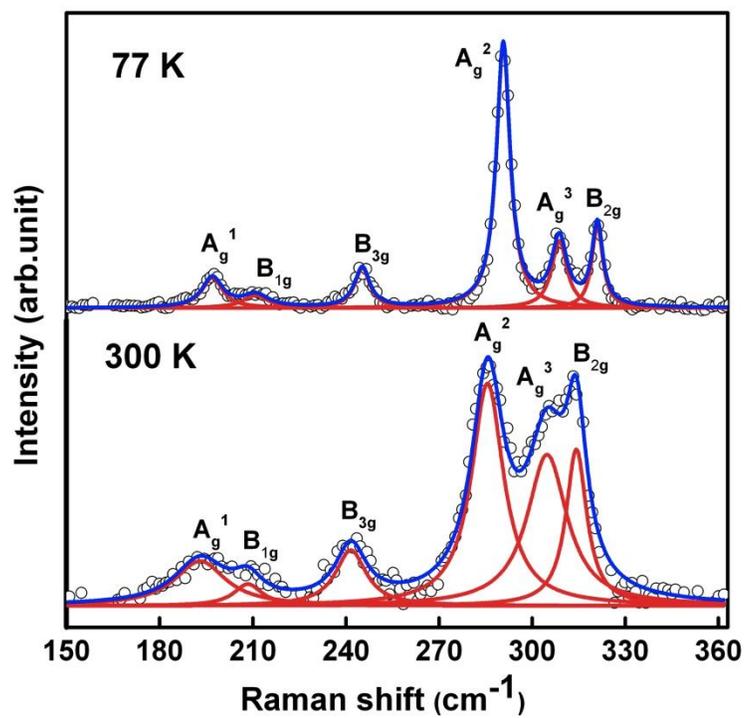

**Fig. 2**.

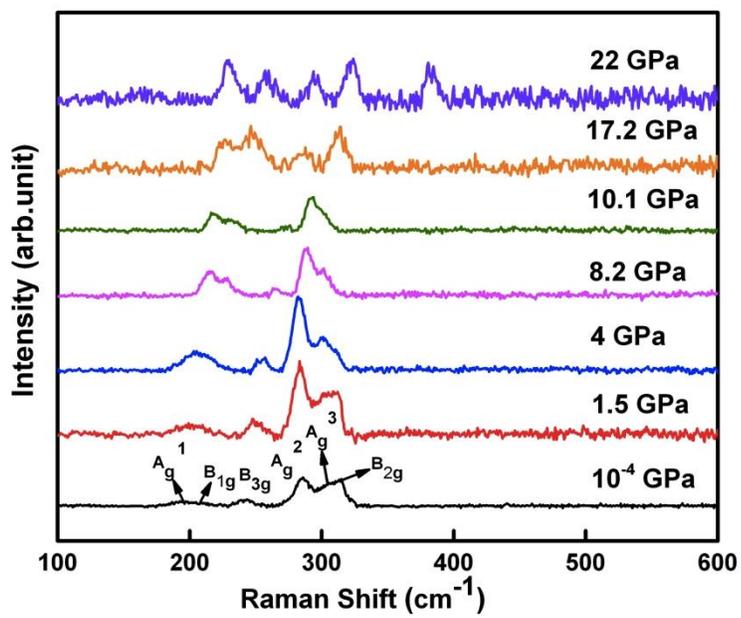

**Fig. 3.**

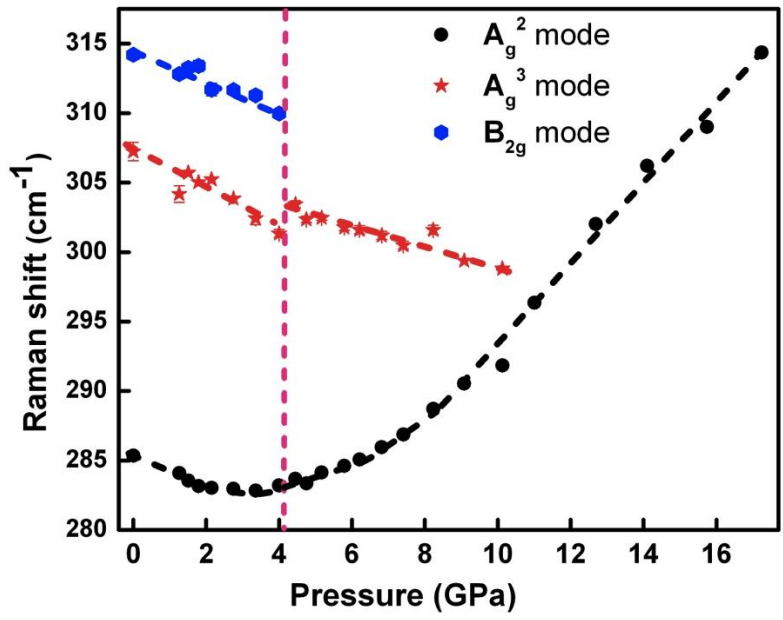

**Fig. 4.**

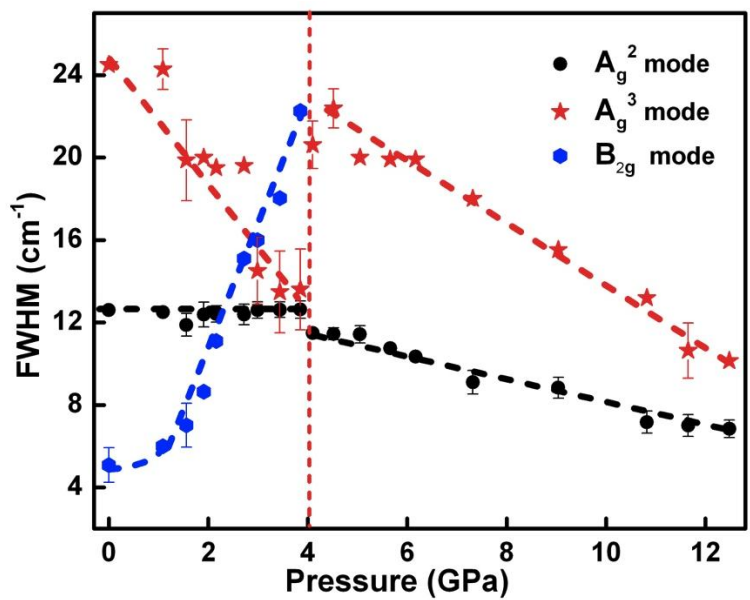

**Fig. 5.**

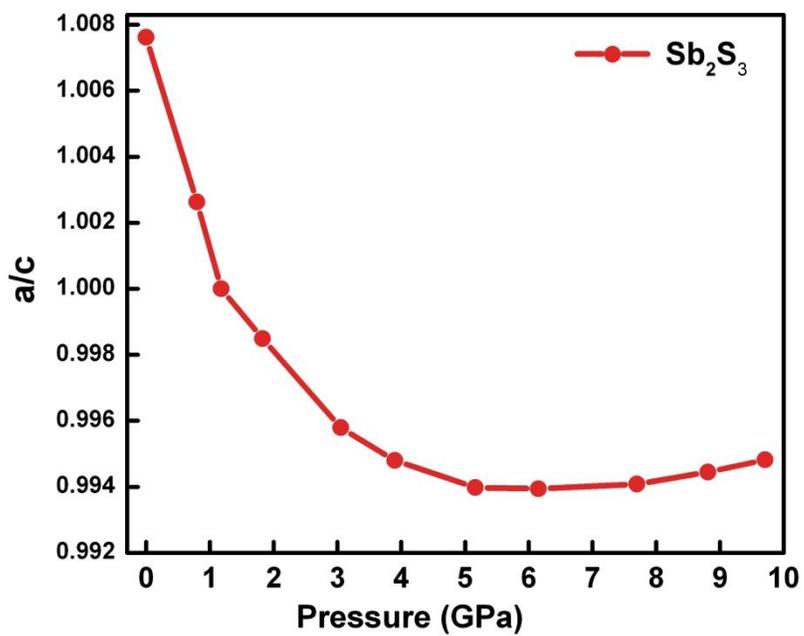

**Fig. 6.**

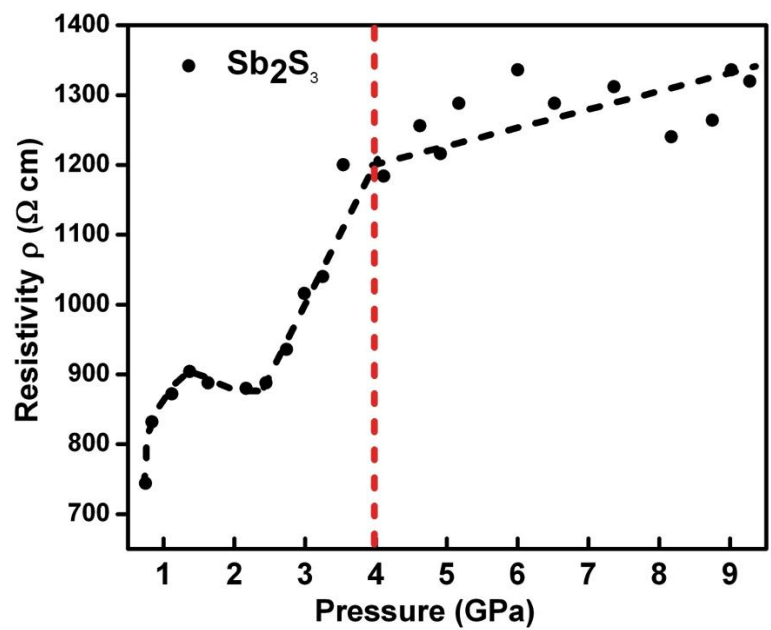